\documentclass[english,british,3p]{elsarticle}
\usepackage[latin9]{inputenc}
\usepackage{float}
\usepackage{amssymb}
\usepackage{graphicx}
\usepackage{babel}
\begin{document}

\title{Basin stability approach for quantifying responses of multistable systems with parameters mismatch}

\author{P. Brzeski{\textsuperscript{1}}, M. Lazarek{\textsuperscript{1}},
T. Kapitaniak{\textsuperscript{1}}, J. Kurths{\textsuperscript{2.3.4.5}},
P. Perlikowski{\textsuperscript{1}}{*}}

\address{\textsuperscript{1}Division of Dynamics, Lodz University of Technology,
90-924 Lodz, Poland}

\address{\textsuperscript{2}
Potsdam Institute for Climate Impact Research, Potsdam 14415,Germany}

\address{\textsuperscript{3}
	Institute of Physics, Humboldt University of Berlin, Berlin 12489,Germany}

\address{\textsuperscript{4}
Department of Control Theory, Nizhny Novgorod State University, Nizhny
Novgorod 603950, Russia}

\address{\textsuperscript{5}
Institute for Complex Systems and Mathematical Biology, University
of Aberdeen, Aberdeen AB24 3UE, UK}

\address{{*}przemyslaw.perlikowski@p.lodz.pl}

\begin{abstract}
In this paper we propose a new method to detect and classify coexisting
solutions in nonlinear systems. We focus on mechanical and structural
systems where we usually avoid multistability for safety and reliability.
We want to be sure that in the given range of parameters and initial
conditions the expected solution is the only possible or at least
has dominant basin of attraction. We propose an algorithm to estimate the
probability of reaching the solution in given (accessible) ranges
of initial conditions and parameters. We use a modified method of basin
stability (Menck et. al., Nature Physics, 9(2) 2013). In our investigation
we examine three different systems: a Duffing oscillator with a tuned
mass absorber, a bilinear impacting oscillator and a beam with attached
rotating pendula. We present the results that prove the usefulness
of the proposed algorithm and highlight its strengths in comparison with
classical analysis of nonlinear systems (analytical solutions, path-following,
basin of attraction ect.). We show that with relatively small
computational effort (comparing to classical analysis) we can predict
the behaviour of the system and select the ranges in parameter's space
where the system behaves in a presumed way. The method can be used
in all types of nonlinear complex systems.
\end{abstract}

\maketitle

\section{Introduction}

In mechanical and structural systems the knowledge of all possible
solutions is crucial for safety and reliability. In devices modelled
by linear ordinary differential equations we can predict the existing
solutions using analytical methods \cite{rao1995mechanical,nayfeh2008linear}.
However, in case of complex, nonlinear systems analytical methods
do not give the full view of system's dynamics \cite{warminski2003approximate,he2004homotopy,belendez2006analytical,nayfeh2011introduction}.
Due to nonlinearity, for the same set of parameters more then one
stable solution may exist \cite{feudel1998dynamical,ChudzikPSK11,Yanchuk2011,gerson2012design,brzeski2012dynamics,menck2013basin,Stability_threshold}.
This phenomenon is called multistability and has been widely investigated
in all types of dynamical systems (mechanical, electrical, biological,
neurobiological, climate and many more). The number of coexisting
solutions strongly depends on the type of nonlinearity, the number of degrees
of freedom and the type of coupling between the subsystems. Hence, usually
the number of solutions vary strongly when values of system's parameters changes. 

As an example, we point out the classical tuned mass absorber
\cite{arnoldfr19955,Cartmell1994173,Alsuwaiyan2002791,fischer2007wind,BVG2008,Ikeda2010,Chung2012,Brzeski2014298}.
This device is well known and widely used to absorb energy and mitigate
unwanted vibrations. However, the best damping ability is achieved
in the neighbourhood of the multistability zone \cite{brzeski2012dynamics}.
Among all coexisting solutions only one mitigates oscillations effectively.
Other solutions may even amplify an amplitude of the base system.
So, it is clear that only by analyzing all possible solutions we can
make the device robust. 

Similarly, in systems with impacts one solution can ensure correct
operation of a machine, while others may lead to damage or destruction
\cite{Brzeski_bells,blazejczyk1998co,de2001basins,qun2003coexisting,de2004controlling}.
The same phenomena is present in multi-degree of freedom systems where
interactions between modes and internal resonances play an important
role \cite{bux1986non,haquang1987non,cartmell1988simultaneous,orlando2013influence}. 

Practically, in nonlinear dynamical systems with more then one degree
of freedom it is impossible to find all existing solutions without
huge effort and using classical methods of analytical and numerical
investigation (path-following, numerical integration, basins of attractions),
especially in cases when we analyse a wider range of system's parameters and we
cannot precisely predict the initial conditions. Moreover, solutions
obtained by integration may have meager basins of attraction and it
could be hard or even impossible to achieve them in reality. That is why
we propose here a new method basing on the idea of basin stability \cite{menck2013basin}.
The classical basin stability method is based on the idea of Bernoulli
trials, i.e., equations of system's motion are integrated $N$ times
for randomly chosen initial conditions (in each trial they are different).
Analyzing the results we asses the stability of each solution. If
there exist only one solution the result of all trials is the same.
But, if more attractors coexist we can estimate the probability of their
occurrence for a chosen set of initial conditions. In mechanical and
structural systems we want to be sure that a presumed solution is stable
and has the dominant basin of attraction in a given range of system's parameters.
Therefore, we build up a basin stability method by drawing values of
system's parameters. We take into account the fact that values of
parameters are measured or estimated with some finite precision and
also that they can slightly vary during normal operation. 

The paper is organized as follows. In Section 2 we introduce simple
models which we use to demonstrate the main idea of our approach.
In the next section we present and describe the proposed method. Section
4 includes numerical examples for systems described in Section 2.
Finally, in Section 5 our conclusions are given.

\section{Model of systems\label{sec:Model-of-systems}}

In this section we present systems that we use to present our method.
Two models are taken from our previous papers \cite{brzeski2012dynamics,czolczynski2012synchronization}
and the third one was described by Pavlovskaia et. al.
\cite{pavlovskaia2010complex}. We deliberately picked models whose
dynamics is well described because we can easily evaluate the correctness
and efficiency of the method we propose.

\subsection{Tuned mass absorber coupled to a Duffing oscillator}

\begin{figure}[H]
\begin{centering}
\includegraphics{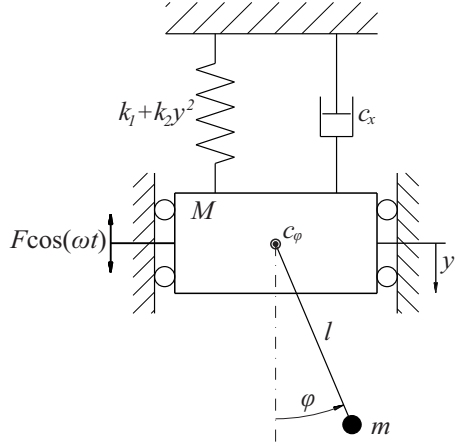}
\par\end{centering}

\caption{The model of the first considered system. Externally forced Duffing
oscillator with attached pendulum (tuned mass absorber). \label{fig:Duffing1}}
\end{figure}

The first example is a system with a Duffing oscillator and a tuned
mass absorber. It was investigated in \cite{brzeski2012dynamics}
and is shown in Figure \ref{fig:Duffing1}. The main body consists
of mass $M$ fixed to the ground with nonlinear spring (hardening
characteristic $k_{1}+k_{2}y^{2}$) and a viscous damper (damping coefficient
$c_{1}$). The main mass is forced externally by a harmonic excitation
with amplitude $F$ and frequency $\omega$. The absorber is modelled
as a mathematical pendulum with length $l$ and mass $m$. A small
viscous damping is present in the pivot of the pendulum. 

The equations of the system's motion are derived in \cite{brzeski2012dynamics},
hence we do not present their dimension form. Based on the following
transformation of coordinates and parameters we reach the dimensionless
form: \foreignlanguage{english}{$\omega_{1}^{2}=\frac{k_{1}}{M+m}$,
$\omega_{2}^{2}=\frac{g}{l}$, $a=\frac{m}{M+m}$, $b=\left(\frac{\omega_{2}}{\omega_{1}}\right)^{2}$,
$\alpha=\frac{k_{2}l^{2}}{(M+m)\omega_{1}^{2}}$, $f=\frac{F}{(M+m)l\omega_{1}^{2}},$
$d_{1}=\frac{c_{x}}{(M+m)\omega_{1}}$, $d_{2}=\frac{c_{\varphi}}{ml^{2}\omega_{2}}$,
$\mu=\frac{\omega}{\omega_{1}}$, $\tau=t\omega_{1}$, $x=\frac{y}{l}$,
$\dot{x}=\frac{\dot{y}}{\omega_{1}l}$, $\ddot{x}=\frac{\ddot{y}}{\omega_{1}^{2}l}$,
$\gamma=\varphi,$ $\dot{\gamma}=\frac{\dot{\varphi}}{\omega_{2}},$
$\gamma=\frac{\ddot{\varphi}}{\omega_{2}^{2}}$.} 

The dimensionless equations are as follows:

\begin{equation}
\begin{array}{c}
\ddot{x}-ab\ddot{\gamma}\sin\gamma-ab\dot{\gamma}^{2}\cos\gamma+x+\alpha x^{3}+d_{1}\dot{x}=f\cos\mu\tau,\\
\\
\ddot{\gamma}-\frac{1}{b}\ddot{x}\sin\gamma+\sin\gamma+d_{2}\dot{\gamma}=0,
\end{array}\label{eq:row bez}
\end{equation}
 where $\mu$ is the frequency of the external forcing and we consider it
as controlling parameter. The dimensionless parameters have the following
values: $f=0.5$, $a=0.091$, $b=3.33$, $\alpha=0.031$, $d_{1}=0.132$
and $d_{2}=0.02$. Both subsystems (Duffing oscillator and the pendulum)
have a linear resonance for $\mu=1.0$.

\subsection{System with impacts}

\begin{figure}
\begin{centering}
\includegraphics{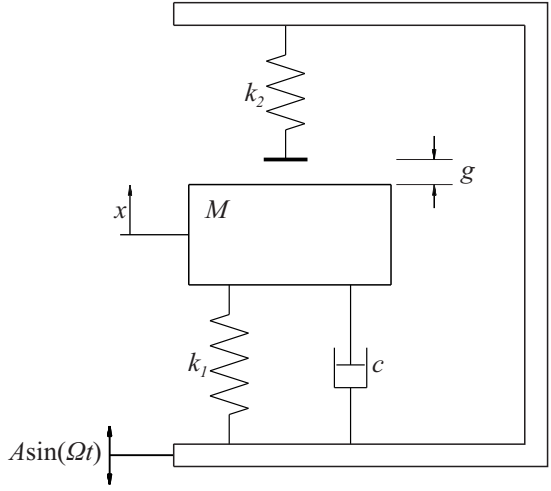}
\par\end{centering}

\caption{The model of the second considered system. Externally forced oscillator with
impacts. \label{fig:Impact}}
\end{figure}

As the next example we analyse a system with impacts \cite{pavlovskaia2010complex}.
It is shown in Figure \ref{fig:Impact} and consists of mass
$M$ suspended by a linear spring with stiffness $k_{1}$ and a viscous
damper with the damping coefficient $c$ to harmonically moving frame.
The frame oscillates with amplitude $A$ and frequency $\Omega$.
When amplitude of mass $M$ motion reaches the value $g$, we observe soft
impacts (spring $k_{2}$ is much stiffer than spring $k_{1}$). 

The dimensionless equation of motion is as follow (for derivation
see \cite{pavlovskaia2010complex}) :

\[
\ddot{x}+2\xi\dot{x}+x+\beta\left(x-e\right)\mathrm{H}\left(x-e\right)=a\omega^{2}\sin\left(\omega\tau\right)
\]

where $x=\frac{y}{y_{0}}$ is the dimensionless vertical displacement
of mass $M$, $\tau=\omega_{n}t$ is the dimensionless time, $\omega_{n}=\frac{k_{1}}{M}$,
$\beta=\frac{k_{2}}{k_{1}}$ the stiffness ratio, $e=\frac{g}{y_{0}}$
 the dimensionless gap between equilibrium of mass $M$ and the
stop suspended on the spring $k_{2}$, $a=\frac{A}{y_{0}}$ and $\omega=\frac{\Omega}{\omega_{n}}$
are dimensionless amplitude and frequency of excitation, $\xi=\frac{c}{2m\omega_{n}}$
is the damping ratio, $y_{0}=1.0\:[\mathrm{mm}]$ and $\mathrm{H}(\cdot)$
 the Heaviside function. In our calculations we take the following
values of system's parameters: $a=0.7$, $\xi=0.01$, $\beta=29$,
$e=1.26$. As a controlling parameter we use the frequency of excitation
$\omega$.

\subsection{Beam with suspended rotating pendula}

\begin{figure}
\begin{centering}
\includegraphics{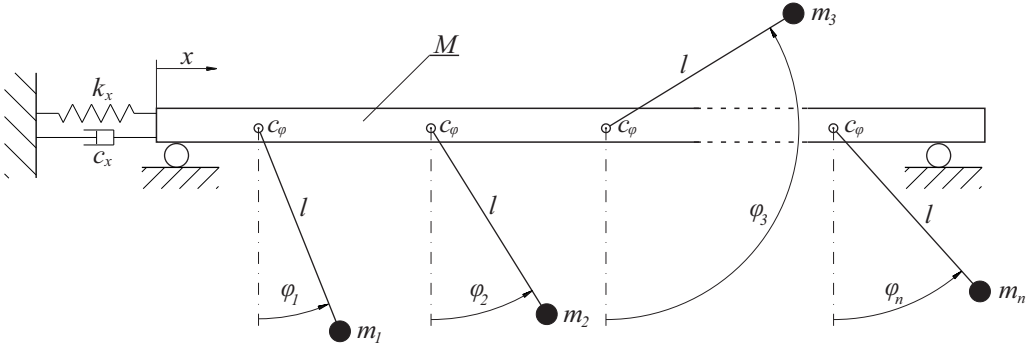}
\par\end{centering}

\caption{The model of the third considered system. Horizontally moving beam
with attached pendulums. \label{fig:Beam_model}}
\end{figure}

The last considered system consists of a beam which can move in
the horizontal direction and $n$ rotating pendula. The beam has the mass
$M$ and supports $n$ rotating, excited pendula. Each pendulum has
the same length $l$ and masses $m_{i}$ $(i=1,\:2,\ldots,\:n)$.
We show the system in Figure \ref{fig:Beam_model} \cite{czolczynski2012synchronization}.
The rotation of the $i$-th pendulum is given by the variable $\varphi_{i}$
and its motion is damped by the viscous friction described by the damping
coefficient $c_{\varphi}$. The forces of inertia of each pendulum
acts on the beam causing its motion in the horizontal direction (described
by the coordinate $x$). The beam is considered as a rigid body, so
we do not consider the elastic waves along it. We describe the phenomena
which take place far below the resonances for longitudinal oscillations
of the beam. The beam is connected to a stationary base by a light
spring with the stiffness coefficient $k_{x}$ and viscous damper with
a damping coefficient $c_{x}$. The pendula are excited by external
torques proportional to their velocities: $N_{0}-\dot{\varphi}_{i}N_{1}$,
where $N_{0}$ and $N_{1}$ are constants. If no other external forces
act on the pendulum, it rotates with the constant velocity $\omega=N_{0}/N_{1}$.
If the system is in a gravitational field (where $g=9.81\:[\mathrm{m/s^{2}}]$
is the acceleration due to gravity), the weight of the pendulum causes
the unevenness of its rotation velocity, i.e., the pendulum slows
down when the centre of its mass goes up and accelerates when the
centre of its mass goes down. 

The system is described by the following set of dimensionless equations:

\begin{equation}
m_{i}l^{2}\ddot{\varphi}_{i}+m_{i}\ddot{x}l\cos{\varphi_{i}}+c_{\varphi}\dot{\varphi}_{i}+m_{i}gl\sin{\varphi_{i}}=N_{0}-\dot{\varphi}_{i}N_{1}\label{eq:Beam1}
\end{equation}

\begin{equation}
\left({M+\sum\limits _{i=1}^{n}{m_{i}}}\right)\ddot{x}+c_{x}\dot{x}+k_{x}x=\sum\limits _{i=1}^{n}{m_{i}l\left({-\ddot{\varphi}_{i}\cos\varphi_{i}+\dot{\varphi}_{i}^{2}\sin\varphi_{i}}\right)}\label{eq:Beam2}
\end{equation}
In our investigation we analyze two cases: a system with two pendula
(where $n=2$ and $i=1,\:2$) and with $20$ pendula ($n=20$ $i=1,\:2,\:...,n$)..
The values of the parameters are as follows: $m_{i}=\frac{2.00}{n}$,
$l=0.25$, $c_{\varphi}=\frac{0.02}{n}$, $N_{0}=5.0$0, $N_{1}=0.50$,
$M=6.00$, $g=9.81$, $c_{x}=\frac{\ln\left(1.5\right)}{\pi}\sqrt{k_{x}\left(M+\sum\limits _{i=1}^{n}{m_{i}}\right)}$and
$k_{x}$ is a controlling parameter. The derivation of the system's equations
can be found in \cite{czolczynski2012synchronization}. We present the
transformation to a dimensionless form in Appendix A.

\section{Methodology}

In \cite{menck2013basin} Authors present a ``basin stability''
method which let us estimate the stability and number of solutions
for given values of system parameters. The idea behind basin stability
is simple, but it is a powerful tool to assess the size of complex
basins of attraction in multidimensional systems. For fixed values
of system's parameters, $N$ sets of random initial conditions are
taken. For each set we check the type of final attractor. Based on
this we calculate the chance to reach a given solution and determine
the distribution of the probability for all coexisting solutions. This
gives us information about the number of stable solutions and the
sizes of their basins of attraction.

We consider the dynamical system $\dot{\mathbf{x}}=f\,(\mathbf{x},\,\omega)$,
where $\mathbf{x}\in\mathtt{\mathbb{R}^{n}}$ and $\omega\in\mathbb{R}$
is the system's parameter. Let ${\cal B\subset\mathbb{\mathtt{\mathbb{R}^{n}}}}$
be a set of all possible initial conditions and ${\cal C}\subset\mathbb{\mathtt{\mathbb{R}}}$
a set of accessible values of system's parameter. Let us assume that an
attractor ${\cal A}$ exists for $\omega\in{\cal C_{A}}\subset{\cal C}$
and has a basin of attraction $\beta({\cal A})$. Assuming random
initial conditions the probability that the system will reach attractor
${\cal A}$ is given by $p\left({\cal A}\right)$. If this probability
is equal to $p\left({\cal A}\right)=1.0$ this means that the considered
solution is the only one in the taken range of initial conditions
and given values of parameters. Otherwise other attractors coexist.
The initial conditions of the system are random from set ${\cal B_{A}}\subset{\cal B}$.
We can consider two possible ways to select this set. 

\begin{description}
  \item[I] The first ensures that set the ${\cal B_{A}}$ includes values of initial 
conditions leading to all possible solutions. This approach is appropriate if 
we want to get a general overview of the system's dynamics. 
  \item[II] In the second approach we use a narrowed set of initial conditions that 
corresponds to practically accessible initial states.\ldots
\end{description}

In our method we chose the second approach because it let us take into 
account constrains imposed on the system and because in engineering we 
usually know or expect the initial state of the system with some finite 
precision. 

In the classical approach of Menck et. al. \cite{menck2013basin} the
values of system's parameters are fixed and do not change during calculations.
The novelty of our method is that we not only draw initial conditions
but also values of some selected parameters of the system. We assume
that the initial conditions and some of the system's parameters are chosen
randomly. Then using $N$ trials of numerical simulations we estimate
the probability that the system will reach a given attractor ${\cal A}$
($p\left({\cal A}\right)$). The idea is to take into consideration
the fact that the values of system's parameters are measured or estimated
with some finite accuracy which is often hard to determine. Moreover
values of parameters can vary during normal operation. Therefore drawing
valus of parameters we can describe how a mismatch in their values influences
the dynamics of the system and estimate the risk of failure. In many
practical applications one is interested in reaching only one presumed
solution ${\cal A}$, and the precise description of other coexisting
attractors is not necessary. We usually want to know the probability
of reaching the expected solution $p\left({\cal A}\right)$ and the chance
that the system behave differently. If $p\left({\cal A}\right)$ is
sufficiently large, we can treat the other attractors as an element
of failure risk. 

In our approach we perform the following steps:
\begin{description}
	\item[I] We pick values of system's parameters from the set
		${\cal C_{A}}\subset{\cal C}$. 
	\item[II]We select the set ${\cal C_{A}}$ so that
		it consists of all practically accessible values of system's parameters
		$\omega$ . This let us ensure that a given solution indeed exists in a practically
		accessible range (taking into account the mismatch in parameters).
	\item[III] We subdivide the set $C_{A}$ in to $m=1,2,\dots M$ equally spaced
		subsets. The subsets ${\cal C}_{A}^{m}$ do not overlap and the relation
		$\bigcup_{m=1\dots M}{\cal C}_{A}^{m}=C_{A}$ is always fulfilled.
	\item[IV] Then for each subset ${\cal C}_{A}^{m}$ we randomly pick $N$ sets
		of initial conditions and value of the considered parameter. For each
		set we check the final attractor of the system. 
	\item[V] After a suficient number
		of trials we calculate the probability of reaching a presumed solution
		or solutions. 
	\item[VI]Finally we describe the relation between the
		value of the system's parameter and the ``basin stability'' of reachable
		solutions.\ldots
\end{description}

In our calculations for each range of parameter values
(subset ${\cal C}_{A}^{m}$) we draw from $N=100$ up to $N=1000$
sets of initial conditions and parameter. The value of $N$ strongly depends
on the complexity of the analysed system. Also the computation time for a
single trial should be adjusted for each system independently
such that it can reach the final attractor. In general, we recommend that
in most cases $N$ should be at least 100.

\section{Numerical results}

\subsection{Tuned mass absorber coupled to a Duffing oscillator}

\begin{figure}[H]
\begin{centering}
\includegraphics{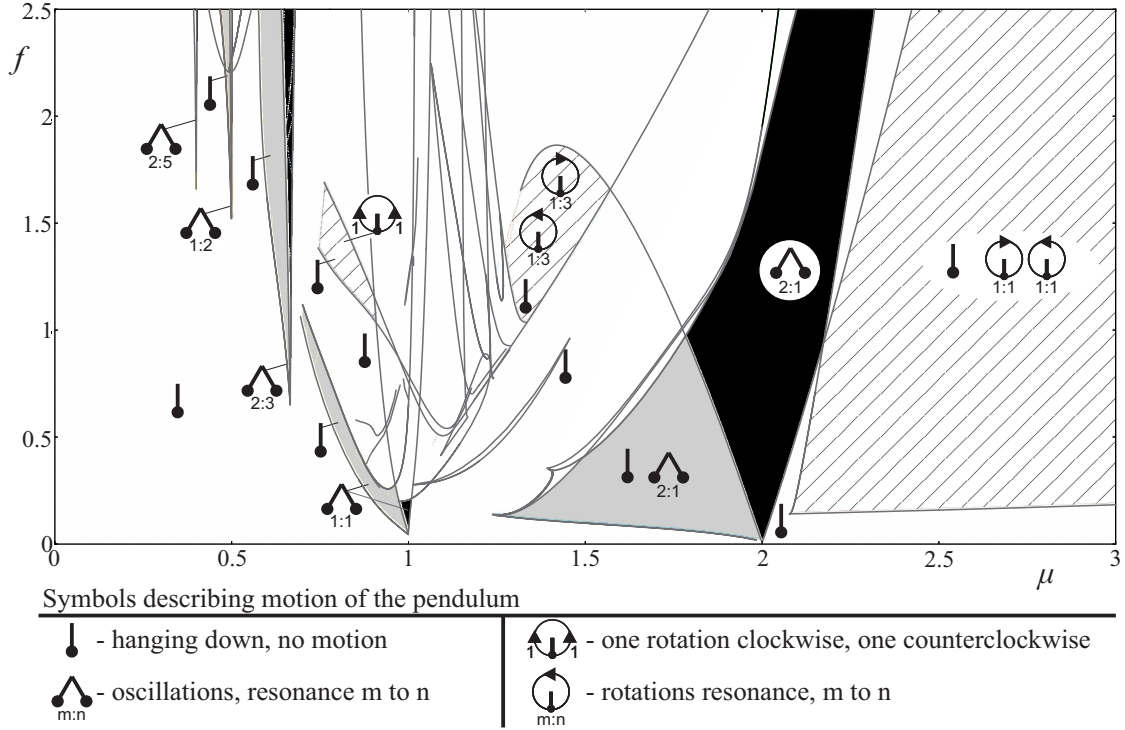}
\par\end{centering}

\caption{Two-parameter bifurcations diagram of the system (1) in the plane $\left(f,\,\mu\right)$
showing periodic oscillations and rotations of the pendulum. Black colour
indicates one attractor, grey colour shows two coexisting attractors
(the same as for black but with a coexisting stable steady state of
the pendulum). In the hatched area we observe the coexistence of stable
rotations and a stable steady state of the pendulum. A detailed analysis
is presented in \cite{brzeski2012dynamics}. \label{fig:Two-paramters_colour}}
\end{figure}

At the beginning we want to recall the results we present in our previous
paper \cite{brzeski2012dynamics}. As a a summary we show Figure \ref{fig:Two-paramters_colour}
with a two dimensional bifurcation diagram obtained by the path-following
method. It gives bifurcations for varying amplitude $f$ and frequency
$\mu$ of the external excitation (see Eq. \ref{eq:row bez}). Lines shown
in the plot correspond to different types of bifurcations (period
doubling, symmetry breaking, Neimark-Sacker and resonance tongues).
We present these lines in one style because the structure is too complex
to follow bifurcation scenarios and we do not need that data (details
are shown in \cite{brzeski2012dynamics}). We mark areas where we
observe the existence of one solution (black colour), or the coexistence of two (grey) and three
(hatched area) stable solutions. The remaining part of the diagram
(white area) corresponds to situations where there are four or more
solutions. Additionally, by white colour we also mark areas where
only the Duffing system is oscillating in 1:1 resonance with the frequency
of excitation and the pendulum is in a stable equilibrium position,
i.e., HDP (hanging down pendulum) state. In this case the dynamics
of the system is reduced to the oscillations of summary mass ($M+m$).

The detailed analysis of system \ref{eq:row bez} is time consuming
and creation of Figure \ref{fig:Two-paramters_colour} was preceded
by complex analysis done with large computational effort. Additionally,
the obtained results give us no information about the size of the basins
of attraction of each solution - which practically means that some
of the solutions may occur only very rarely in the real system (i.e. due to not accessible
initial conditions). Nevertheless, such analysis gives us an in-depth
knowledge about the bifurcation structure of the system. As we can see, the
range where less then three solutions exist is rather small, especially
for $\mu<2.0$. To illustrate our method of analysis, we focus on three
solutions: $2:1$ oscillating resonance, HDP and $1:1$ rotating resonance
assuming that only they have some practical meaning.

\begin{figure}[H]
\begin{centering}
\includegraphics{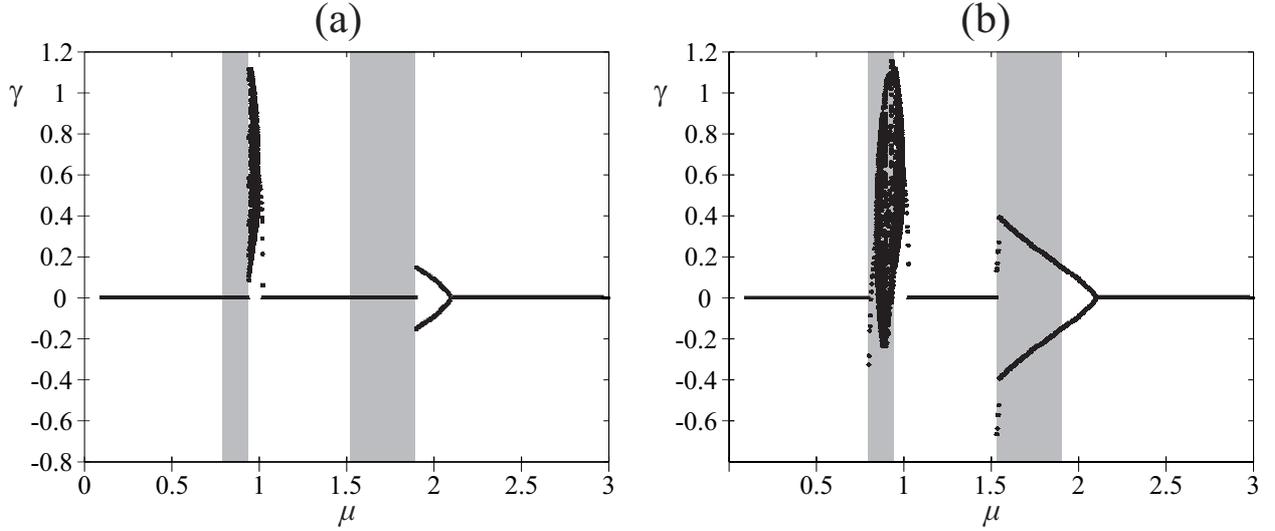}
\par\end{centering}

\caption{Bifurcation diagram showing the behaviour of the pendulum suspended
on the Duffing oscillator. For subplot (a) the value of the bifurcation parameter
$\mu$ was increased, while for subplot (b) we decreased the value of $\mu$.
Gray rectangles mark the range of the bifurcation parameter $\mu$ for
which different attractors coexist. A detailed analysis is presented
in \cite{brzeski2012dynamics}. \label{fig:tma_bif}}
\end{figure}

To show our results obtained with integration, we compute bifurcation diagrams
for $f=0.5$ in the range $\mu\in[0.1,\:3.0]$ (see Figure \ref{fig:tma_bif}).
In Figure \ref{fig:tma_bif}(a) we increase $\mu$ from $0.1$ to
$3.0$ and in Figure \ref{fig:tma_bif}(b) we decrease $\mu$ from $3.0$
to $0.1$. As the initial conditions we take the equilibrium position
($x_{0}=\dot{x}_{0}=0.0$ and $\gamma_{0}=\dot{\gamma}_{0}=0.0$).
In both panels we plot the amplitude of the pendulum $\gamma$. Ranges where the 
diagrams differ we mark by grey rectangles. It is easy to see that there
are two dominating solutions: HDP and $2:1$ internal resonance. Near
$\mu=1.0$ we observe a narrow range of $1:1$ and $9:9$ resonances
and chaotic motion (for details see Figure 6 in \cite{brzeski2012dynamics}).
Based on previous results we know that we detected all solutions existing
in the considered range, however we do not have information about the size of
their basins of attraction and coexistence. Hence the analysis with the 
proposed method should give us new important information about the system's
dynamics. Contrary to the bifurcation diagram obtained by path-following
in Figure \ref{fig:tma_bif}, we do not observe rotating solutions
(the other set of initial conditions should be taken).

\begin{figure}[H]
\begin{centering}
\includegraphics{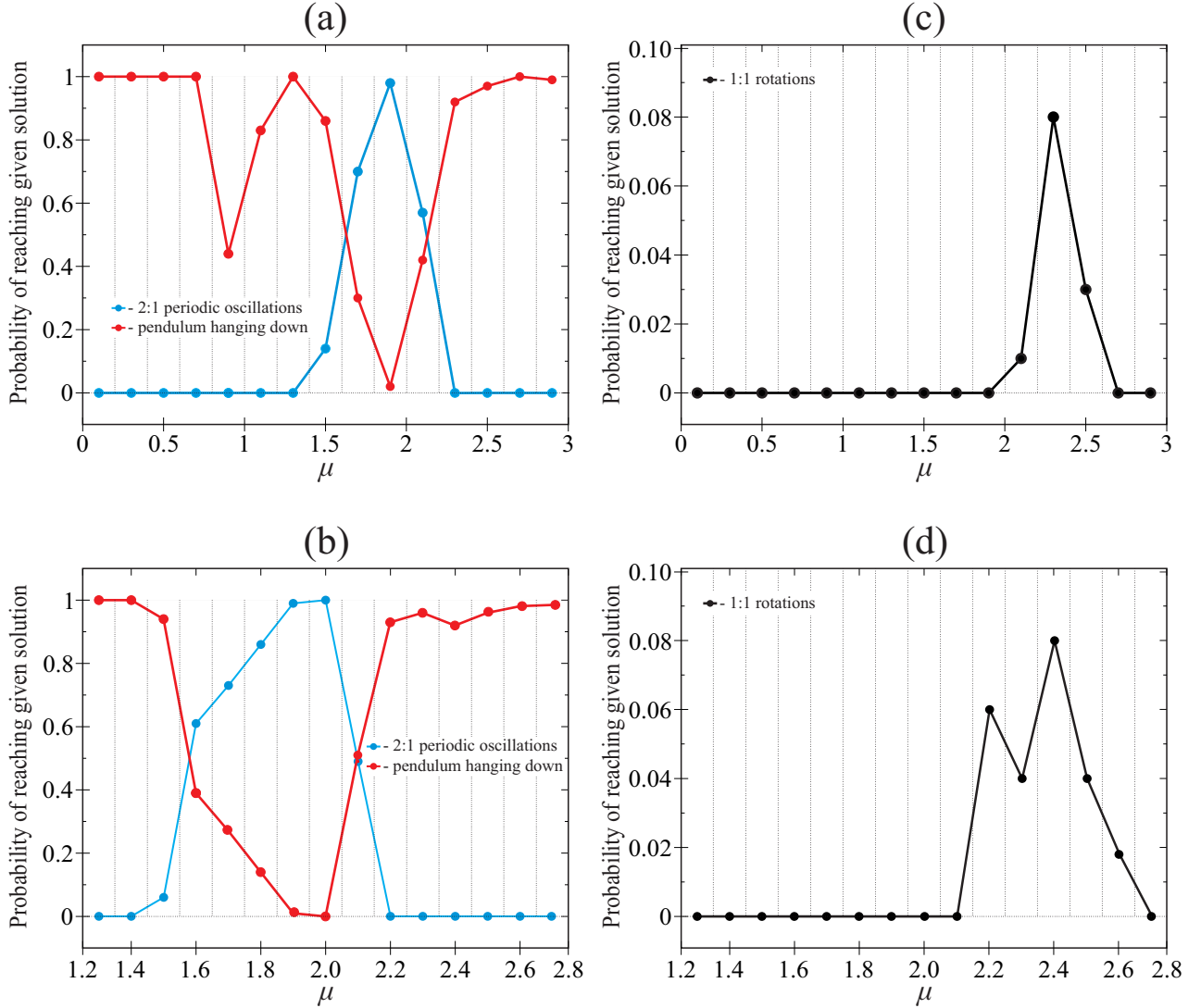}
\par\end{centering}

\caption{Probability of reaching given solutions in (1) system with tuned mass
absorber. Subplots (a,b) present solutions with $2:1$ periodic oscillations
(blue) and without motion of the pendulum (red). Subplots (c,d) present the
probability of reaching $1:1$ rotations (black). (Please note that
in both cases (a,b) and (c,d) the initial conditions and parameter
are somehow random, hence the results may slightly differ). \label{fig:Duff_prob}}
\end{figure}

In Figure \ref{fig:Duff_prob} we show the probability of reaching the
three aforementioned solutions obtained using the proposed method.
The initial conditions are random numbers drawn from the following
ranges: $x_{0}\in[-2,\:2]$, $\dot{x}_{0}\in[-2,\:2]$, $\gamma_{0}\in[-\pi,\:\pi]$
and $\dot{\gamma}_{0}\in[-2.0,\:2.0]$ (ranges there selected basing
on the results from \cite{brzeski2012dynamics}). The frequency of excitation
is within a range $\mu\in[0,\:3.0]$ (Figure \ref{fig:Duff_prob}(a,c)
), then we refine it to $\mu\in[1.25,\:2.75]$ (Figure \ref{fig:Duff_prob}(b,d)
). In both cases we take $15$ equally spaced subsets of $\mu$ and
in each subset we calculate the probability of reaching a given solution.
For each subset we calculate $100$ trials each time drawing initial
conditions of the system and a value of $\mu$ from the appropriate
range. Then we plot the dot in the middle of the subset which indicate
the probability of reaching a given solution in each considered range.
Lines that connect the dots are shown just to ephasize the tendency. For
each range we take $N=1000$ because we want to estimate the probability
of a solution with small a basin of stability (1:1 rotating periodic solution).

As we can see in Figure \ref{fig:Two-paramters_colour}, the $2:1$
resonance solution exists in the area marked by black colour around
$\mu=2.0$ and coexists with HDP in the neighbouring grey zone. In Figure
\ref{fig:Duff_prob} we mark the probability of reaching the $2:1$ resonance
using blue dots. As we expected, for $\mu<1.4$ and $\mu>2.2$ the solution
does not exist. In the range $\mu\in[1.4,\:2.2]$ the maximum value of
probability $p(2:1)=0.971$ is reached in the subset $\mu\in[1.8,\:2.0]$
and outside that range the probability decreases. To check if we can
reach $p(2:1)=1.0$, we decrease the range of parameter's values to
$\mu\in[1.25,\:2.75]$ and the size of subset to $\Delta\mu=0.1$
(we still have 15 equally spaced subsets). The results are shown in
Figure \ref{fig:Duff_prob}(b) similarly in blue colour. In the range
$\mu\in[1.95,\:2.05]$ the probability $p(2:1)$ is equal to unity
and in the range $\mu\in[1.85,\:1.95]$ it is slightly smaller $p(2:1)=0.992$.
Hence, for both subsets we can be nearly sure that the system reaches the $2:1$
solution. This gives us indication of how precise we have to set the
value of $\mu$ to be sure that the system will behave in a presumed
way.

A similar analysis is performed for HDP. The values of probability
is indicated by the red dots. As one can see for $\mu<0.8$, $\mu\in[1.2,\:1.4]$
and $\mu\in[2.6,\:2.8]$, the HDP is the only existing solution. The
rapid decrease close to $\mu\approx1.0$ indicates the $1:1$ resonance
and the presence of other coexisting solutions in this range (see
\cite{brzeski2012dynamics}). In the range $\mu\in[1.2,\:1.4]$ the probability
$p(\mathrm{HDP})=1.0$ which corresponds to a border between solutions
born from $1:1$ and $2:1$ resonance. Hence, up to $\mu=2.0$ the
probability of the HDP solution is a mirror refection of $p(2:1)$. The
same tendency is observed in the narrowed range as presented in Figure
\ref{fig:Duff_prob}(b). Finally, for $\mu>2.0$ the third considered
solution comes in and we start to observe an increase of probability
of the rotating solution $S(\mu,\:\mathrm{HDP})$ as shown in Figure \ref{fig:Duff_prob}(c).
However, the chance of reaching the rotating solution remains small and never
exceeds $p(\mathrm{1:1})=8\times10^{-3}$. We also plot the probability
of reaching the rotating solution in the narrower range of $\mu$ in Figure
\ref{fig:Duff_prob}(d). The probability is similar to the one presented
in Figure \ref{fig:Duff_prob}(c) - it is low and does not exceed $p(\mathrm{1:1})=8\times10^{-3}$.
Note that the results presented in Figure \ref{fig:Duff_prob}(a,b) and
Figure \ref{fig:Duff_prob}(c,d) are computed for different sets of
random initial conditions and parameter values; hence the obtained
probability can be slightly different. 

\subsection{System with impacts}

\begin{figure}
\begin{centering}
\includegraphics{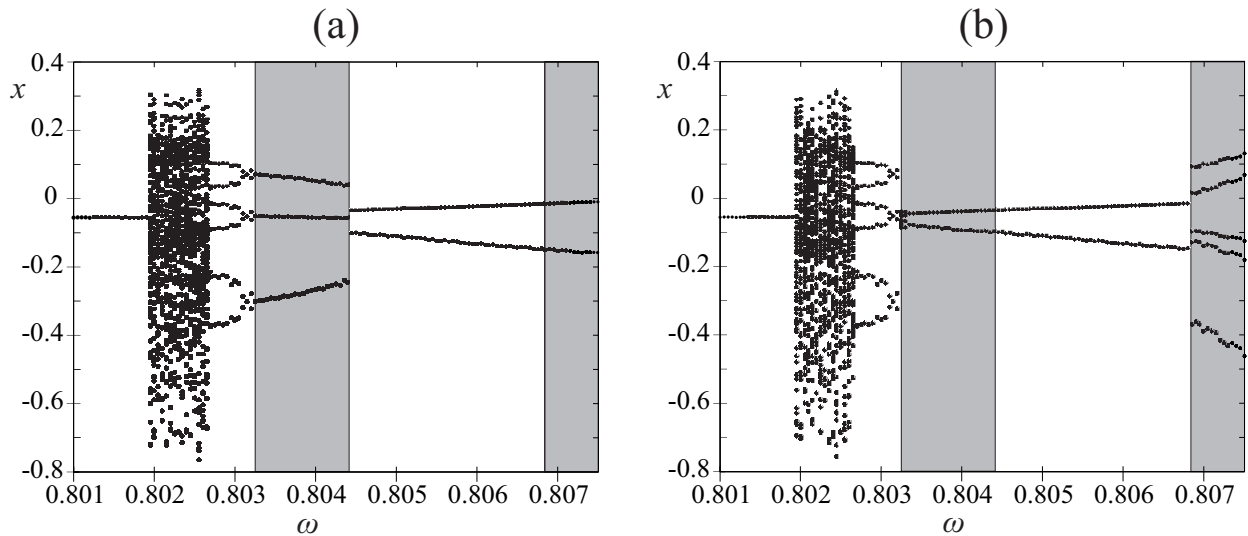}
\par\end{centering}

\caption{Bifurcation diagram showing the behaviour of impacting oscillator (2).
For subplot (a) the value of the bifurcation parameter $\omega$ was increased
while for subplot (b) we decreased the value of $\omega$. Grey rectangles
mark the range of the bifurcation parameter $\omega$ for which different
attractors coexist. Further analysis can be found in \cite{pavlovskaia2010complex}.
\label{fig:ImpactBif}}
\end{figure}

In this subsection we present our analysis of different periodic solutions
in the system with impacts. A discontinuity usually increases the number
of coexisting solutions. Hence, in the considered system we observe
a large number of different stable orbits and their classification
is necessary. In Figure \ref{fig:ImpactBif} we show two bifurcation
diagrams with $\omega$ as controlling parameter. Both of them start
with initial conditions $x_{0}=0.0$ and $\dot{x}_{0}=0.0$. In panel
(a) we increase $\omega$ from $0.801$ to $0.8075$; while in panel
(b) we decrease $\omega$ in the same range. We select the range of
$\omega$ basing on the results presented in \cite{pavlovskaia2010complex}.
As one can see, both diagrams differ in two zones marked by grey colour.
Hence, we observe a coexistence of different solutions, i.e., in the range
$\omega\in[0.8033,\:0.8044]$ solutions with period-3 and -2 are present,
while in the range $\omega\in[0.8068,\:0.8075]$ we detected solutions
with period-2 and -5. As presented in \cite{pavlovskaia2010complex}
some solutions appear from a saddle-node bifurcation and we are not
able to detect them with the classical bifurcation diagram. The proposed
method solves this problem and shows all existing solutions in
the considered range of excitation frequency. 

\begin{figure}
\begin{centering}
\includegraphics{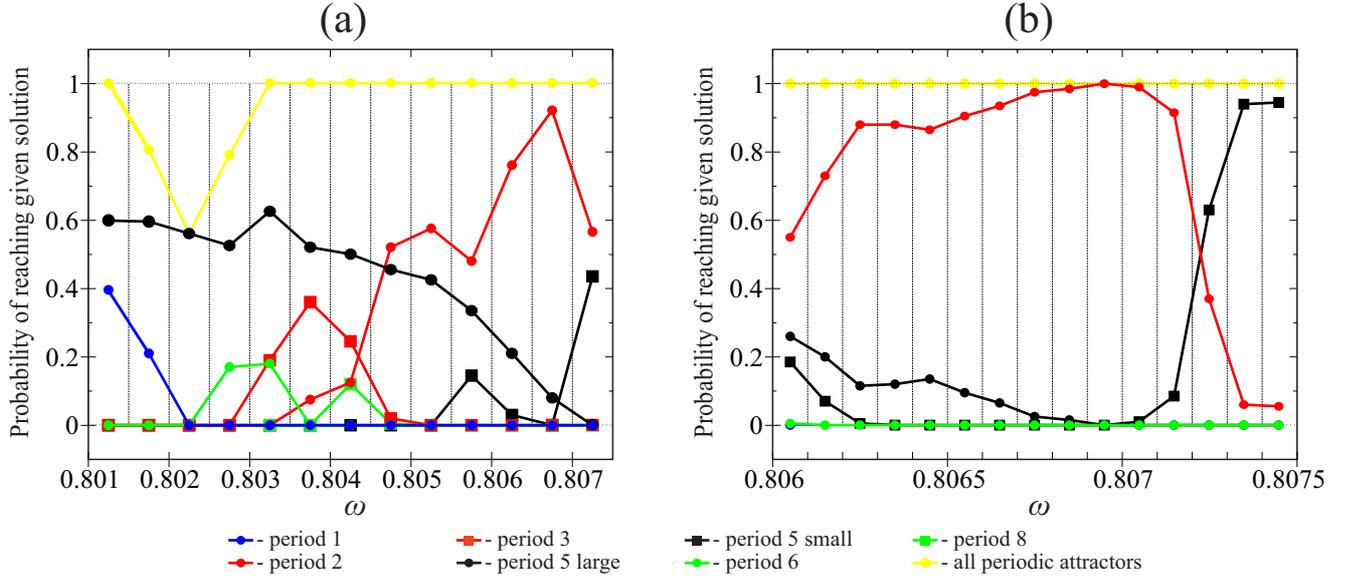}
\par\end{centering}

\caption{Probability of reaching given solutions in the impacting system. Subplots
(a,b) present different periodic solutions and the summary probability
of reaching any periodic solution. In Subplot (a) we analyze $\omega\in[0.801,\:0.8075]$
with the step $\Delta\omega=0.0005$, and in subplot (b) we narrow
the range $\omega\in[0.806,\:0.8075]$ and decrease the step size
$\Delta\omega=0.0001$. (Please note that in cases (a) and (b) the
initial conditions and parameter are somehow random, hence the results
may slightly differ).\label{fig:Impact_prob}}
\end{figure}

We focus on periodic solutions with periods that are not longer than
eight periods of excitation. We observe periodic solutions with higher
periods in the narrow range of $\omega$ but the probability that
they will occur is very small and we can neglect them. All non-periodic
solutions are chaotic (quasiperiodic solutions are not present in
this system). The results of our calculations are shown in Figure
\ref{fig:Impact_prob}(a,b). We take initial conditions from the following
ranges $x_{0}\in[-2,\:2]$, $\dot{x}_{0}\in[-2,\:2]$. The controlling
parameter $\omega$ is changed from $0.801$ to $0.8075$ with step
$\Delta\omega=0.0005$ in Figure \ref{fig:Impact_prob}(a) and from
$0.806$ to $0.8075$ with the step $\Delta\omega=0.0001$ in Figure \ref{fig:Impact_prob}(b)
(in each subrange of excitation's frequency we pick the exact value of
$\omega$ randomly from this subset). The probability of periodic
solutions is plotted by lines with different colours and markers.
We detect the following solutions: period-1, -2, -3, -5 (two different
attractors with large and small amplitude), -6 and -8. The dot lines
indicate the sum of all periodic solutions' probability (also with
period higher then eight). Hence, when its value is below $1$, chaotic solution exist. Dots are drawn for mean value i.e, middle
of the subset. For each range we take $N=200$ and we increase the calculation
time because the transient time is sufficiently larger than in the
previous example due to the piecewise smooth characteristic of spring's
stiffness.

As we can see, the chance of reaching a given solution strongly depends
on $\omega$. Hence, in the sense of basin stability we can say that
stability of solutions rely upon the $\omega$ value. In Figure \ref{fig:Impact_prob}(a)
the probability of a single solution is always smaller than one. Nevertheless,
we observe two dominant solutions: period-5 with large amplitude in
the first half of the considered $\omega$ range and period-2 in the second
half of the range. The maximum registered value of probability is $p(\mathrm{period-2})=0.92$
and it refers to the period-2 solution for $\omega\approx0.80675$. To
check if we can achieve even higher probability we analyse a narrower
range of $\omega$ and decrease the step (from $\Delta\omega=0.0005$
to $\Delta\omega=0.0001$). In Figure \ref{fig:Impact_prob}(b) we
see that in range $\omega\in[0.8069,\:0.807]$ the probability of reaching
the period-2 solution is equal to $1$. Hence, in the sense of basin stability
it is the only stable solution. Also in the range $\omega\in[0.8065,\:0.8072]$
the probability of reaching this solution is higher then $0.9$ and
we can say that its basin of attraction is strongly dominant. 

Other periodic solutions presented in Figure \ref{fig:Impact_prob}(a)
are: period-1 is present in the range $\omega\in[0.801,\:0.8025]$ with the
highest probability $p(\mathrm{period-1})=0.4$, period-3 exists in the
range $\omega\in[0.803,\:0.805]$ with the maximum probability $p(\mathrm{period-3})=0.36$,
period-2 is observed in two ranges $\omega\in[0.8025,\:0.8035]$ and
$\omega\in[0.804,\:0.8045]$ with the highest probability equal to $0.18$
and $0.12$ respectively. Solution with period-5 (small amplitude's
attractor) exists also in two ranges $\omega\in[0.8055,\:0.8065]$
and $\omega\in[0.807,\:0.8075]$ with the highest probability equal to
$0.14$ and $0.4$3 respectively.

\subsection{Beam with suspended rotating pendula}

The third considered system consists of a beam that can move horizontally
with two ($n=2$) or twenty ($n=20$) pendula suspended on it. As
a control parameter we use $k_{x}$ which describes the stiffness
of the beam's support. For the considered range of $k_{x}\in[100,\,5000]$
two stable periodic attractors exist in that system. One corresponds
to complete synchronization of the rotating pendula. The second one is called anti-phase synchronization and refers to the state when
the pendula rotate in the same direction but are shifted in phase by $\pi$.

\begin{figure}
\begin{centering}
\includegraphics{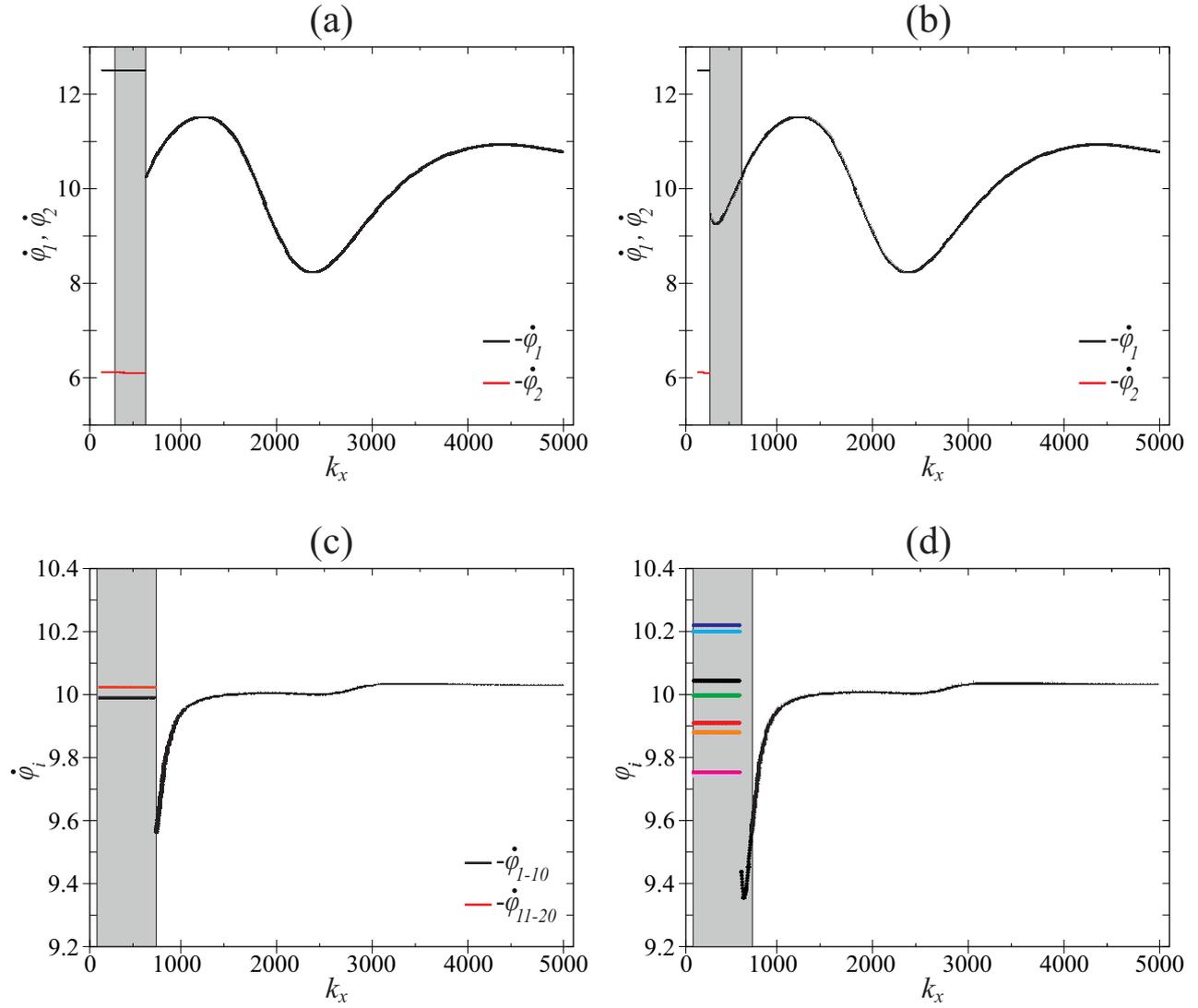}
\par\end{centering}

\caption{Bifurctaion diagram showing the behaviour of two (a,b) and twenty
(c,d) pendula suspended on the moving beam. For subplots (a,c) the value
of the bifurcation parameter $k_{x}$ was increased, while for subplots
(b,d) we decreased the value of $k_{x}$. Grey rectangles mark the
ranges of the bifurcation parameter $k_{x}$ for which different attractors
coexist. Further analysis of number of solutions can be found in \cite{czolczynski2012synchronization}.
\label{fig:pendulaBif}}
\end{figure}

In Figure \ref{fig:pendulaBif} we show four bifurcation diagrams
with $k_{x}$ as the controlling parameter and a Pioncare map of rotational
speed of the pendula. The subplots (a,b) refer to the system with two
pendula ($n=2$). We start with zero initial conditions: $x_{0}=0.0$,
$\dot{x}_{0}=0.0$, $\varphi_{10}=0.0$, $\dot{\varphi}_{10}=0.0$,
$\varphi_{20}=0.0$, $\dot{\varphi}_{20}=0.0$ and take $k_{x}\in[100,\,5000]$.
The parameter $k_{x}$ is increasing in subplot (a) and decreasing in
(b). We see that in the range marked by grey rectangle both complete
and anti-phase synchronization coexist. In subplots (c,d) we present
results for twenty pendula ($n=20$). We start the integration from initial
conditions that refer to anti-phase synchronization (two clusters
of 10 pendula shifted by $\pi$) i.e. $x_{0}=0.1$, $\dot{x}_{0}=0.00057$,
$\varphi_{k0}=0.0$, $\dot{\varphi}_{k0}=9.81$, $\varphi_{j0}=3.09$,
$\dot{\varphi}_{j0}=9.784$ where: $k=1,2,\ldots10$ and $j=11,12,\ldots20$.
The value of $k_{x}$ is increasing in subplot (c) and decreasing in (d).
Similarly as in the two pendula case, we observe the region ($k_{x}\in[100,\,750]$)
where two solutions coexist: anti-phase synchronization and non-synchronous
state. To further analyse multistability in that system we use 
proposed method.

\begin{figure}[H]
\begin{centering}
\includegraphics{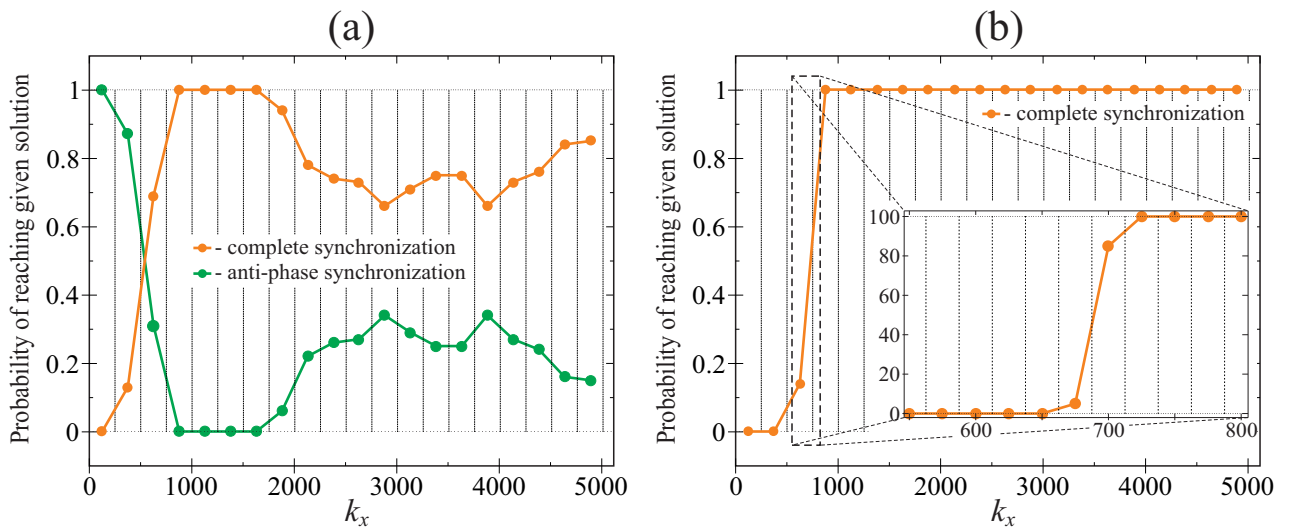}
\par\end{centering}

\caption{Probability of reaching given solutions in the system with rotating
pendula. Subplot (a) refers to the case with two pendula and (b) with
twenty pendula. (Please note that on plot (b) and its magnification
the initial conditions and parameter are somehow random, hence the
results may slightly differ). \label{fig:Pendula_prob}}
\end{figure}

In Figure \ref{fig:Pendula_prob} we present how the probability of
reaching a given solution depends on the parameter $k_{x}$ . In subplot
(a) we show the results for the system with 2 pendula, while in subplot
(b) results obtained for the system with 20 pendula suspended
on the beam are given. In both cases we consider $k_{x}\in[0,\:5000]$ and assume
the following ranges of initial conditions: $x_{0}\in[-0.15,\:0.15]$,
$\dot{x}_{0}\in[-0.1,\:0.1]$, $\varphi_{i0}\in[-\pi,\:\pi]$, $\varphi_{20}\in[-\pi,\:\pi]$,
$\dot{\varphi}_{10}\in[-3.0,\:3.0]$ and $\dot{\varphi}_{20}\in[-3.0,\:3.0]$
in Figure \ref{fig:Pendula_prob}(a) and $x_{0}\in[-0.15,\:0.15]$,
$\dot{x}_{0}\in[-0.1,\:0.1]$, $\varphi_{i0}\in[-\pi,\:\pi]$, $\dot{\varphi}_{i0}\in[-\pi,\:\pi]$
where $i=1\dots\:20$ in Figure \ref{fig:Pendula_prob}(b). We take
$20$ subsets of parameter $k_{x}$ values with the step equal to
$\Delta k_{x}=250$ and mark their borders with vertical lines. For
each set we run $N=100$ simulations; each one with random initial
conditions and $k_{x}$value drawn from the respective subset. Then,
we estimate the probability of reaching given solution. The dots in
Figure \ref{fig:Pendula_prob} indicate the probability of reaching a
given solution in the considered range (dots are drawn for mean value,
i.e, middle of subset). Contrary to both already presented systems,
this one has a much larger dimension of phase space (six and forty two),
hence we decide to decrease number of the trials to $N=100$ in order
to minimise the time of calculations. 

In Figure \ref{fig:Pendula_prob}(a) we show the results for 2 pendula.
When $k_{x}\in[0,\:250]$ only anti-phase synchronization is possible.
Then, with the increase of $k_{x}$ we observe a sudden change
in the probability and for $k_{x}\in[750,\:1750]$ only complete synchronization
exists. For $k_{x}>2000$ a probability of reaching both solutions fluctuates
around $p(\mathrm{complete})=0.7$ for complete and $p(\mathrm{\mathrm{anti-phas}e})=0.3$
for anti-phase synchronization. Further increase of $k_{x}$
does not introduce any significant changes. 

In Figure \ref{fig:Pendula_prob}(b) we show the results for twenty
pendula. For $k_{x}\in[0,\:250]$ the system reaches solutions different
from the two analysed (usually chaotic). Then, the probability of reaching
complete synchronization drastically increases and for $k_{x}\in[750,\:5000]$
it is equal to $p(complete)=1.0$ which means that the pendula always
synchronize completely. We also present the magnification of the plot
where we see that in fact for $k_{x}\in[715,\:5000]$ we will always
observe complete synchronization of the pendula. Please note that
for calculating both plots we use random initial conditions and $k_{x}$
value hence, the results for a narrower range may differ. Anti-phase
synchronization was never achieved with randomly chosen initial conditions.
This means that even though this solution is stable for $k_{x}\in[100,\:750]$
(see Figure \ref{fig:pendulaBif}(c)) it has a much smaller basin of attraction
and is extremely hard to obtain in reality. The results presented in Figure\ref{fig:Pendula_prob}
prove that by proper tuning of the parameter $k_{x}$ we can control the
systems behaviour even if we can only fix the $k_{x}$ value with finite
precision. 

\section{Conclusions}

In this paper we propose a new method of detection of solutions' in 
non-linear mechanical or structural systems. The method allows
to get a general view of the system's dynamics and estimate the risk that
the system will behave behave differently than assumed. To achieve
this goal we extend the method of basin stability \cite{menck2013basin}.
We build up the classical algorithm and draw not only initial conditions
but also values of system's parameters. We take this into account
because the identification of parameters' values is quite often not
very precise. Moreover values of parameters often slowly vary during
operation. Whereas in practical applications we usually need certainty
that the presumed solution is stable and its basin of stability is large
enough to ensure its robustness. Hence, there is a need to describe
how small changes of parameters' values influence the behaviour of
the system. Our method provides such a description and allows us to estimate
the required accuracy of parameters values and the risk of unwanted
phenomena. Moreover it is relatively time efficient and does not require
high computational power.

We show three examples, each for a different class of systems: a tuned
mass absorber, a piecewise smooth oscillator and a multi-degree of freedom
system. Using the proposed method we can estimate the number of existing
solutions, classify them and predict their probability of appearance.
Nevertheless, in many cases it is not necessary to distinct all solutions
existing in a system but it is enough to focus on an expected solution,
while usually other periodic, quasi-periodic and chaotic solutions
are classified as undesirable. Such a strategy simplifies the analysis and
reduces the computational effort. We can focus only on probable solutions
and reduce the number of trials omitting a precise description of solutions
with low probability. 

The proposed method is robust and can be used not only for mechanical
and structural systems but also for any system given by differential
equations where the knowledge about existing solutions is crucial.

\section*{Acknowledgement}

This work is funded by the National Science Center Poland based on
the decision number DEC-2015/16/T/ST8/00516. PB
is supported by the Foundation for Polish Science (FNP).

\section*{Appendix A}

The motion of the system presented in Figure \ref{fig:Beam_model}
is described by the following set of two second order ODEs:

\begin{equation}
m_{iD}l_{D}^{2}\ddot{\varphi'}_{i}+m_{iD}\ddot{x'}l_{D}\cos{\varphi'_{i}}+c_{\varphi D}\dot{\varphi'}_{i}+m_{iD}g_{D}l_{D}\sin{\varphi'_{i}}=N_{0D}-\dot{\varphi'}_{i}N_{1D}\label{eq:Beam1-1-1}
\end{equation}

\begin{equation}
\left({M_{D}+\sum\limits _{i=1}^{n}{m_{iD}}}\right)\ddot{x'}+c_{xD}\dot{x'}+k_{xD}x'=\sum\limits _{i=1}^{n}{m_{iD}l_{D}\left({-\ddot{\varphi'}_{i}\cos\varphi'_{i}+\dot{\varphi'}_{i}^{2}\sin\varphi'_{i}}\right)}\label{eq:Beam2-1-1}
\end{equation}

The values of parameters and their dimensions are as follow: $m_{iD}=\frac{2.00}{n}\,[kg]$,
$l_{D}=0.25\,[m]$, $c_{\varphi D}=\frac{0.02}{n}\,[Nms]$, $N_{0D}=5.00\,[Nm]$,
$N_{1D}=0.50\,[Nms]$, $M_{D}=6.00\,[kg]$, $g_{D}=9.81\,[\frac{m}{s^{2}}]$,
$c_{x_{D}}=\frac{\ln\left(1.5\right)}{\pi}\sqrt{k_{x}\left(M+\sum\limits _{i=1}^{n}{m_{i}}\right)}\,[\frac{Ns}{m}]$
and $k_{xD}\,[\frac{N}{m}]$ is controlling parameter. The derivation
of the above equations can be found in \cite{czolczynski2012synchronization}.
\foreignlanguage{english}{We perform a transformation to a dimensionless
form in a way that enables us to hold parameters' values. It is because
we want to present new results in a way that thay can be easily compared
to results of the investigation presented in }\cite{czolczynski2012synchronization}\foreignlanguage{english}{.
We introduce dimensionless time $\tau=t\omega_{0}$, where $\omega_{0}=1\,\mathrm{[Hz]}$,
and unit parameters $m_{0}=1.0\,[kg]$, }$l_{0}=1.0\,[m]$\foreignlanguage{english}{
and reach the dimensionless equations:}

\begin{equation}
m_{i}l^{2}\ddot{\varphi}_{i}+m_{i}\ddot{x}l\cos{\varphi_{i}}+c_{\varphi}\dot{\varphi}_{i}+m_{i}gl\sin{\varphi_{i}}=N_{0}-\dot{\varphi}_{i}N_{1}\label{eq:Beam1-1}
\end{equation}

\begin{equation}
\left({M+\sum\limits _{i=1}^{n}{m_{i}}}\right)\ddot{x}+c_{x}\dot{x}+k_{x}x=\sum\limits _{i=1}^{n}{m_{i}l\left({-\ddot{\varphi}_{i}\cos\varphi_{i}+\dot{\varphi}_{i}^{2}\sin\varphi_{i}}\right)}\label{eq:Beam2-1}
\end{equation}

\selectlanguage{english}%
where: $x=\frac{x'}{l_{0}}$, $\dot{x}=\frac{\dot{x'}}{l_{0}\omega_{0}}$,
$\ddot{x}=\frac{\ddot{x'}}{l_{0}\omega_{0}^{2}}$, $\varphi_{i}=\varphi'_{i}$,
$\dot{\varphi}_{i}=\frac{\dot{\varphi'}_{i}}{\omega_{0}}$, $\ddot{\varphi}_{i}=\frac{\ddot{\varphi'}_{i}}{\omega_{0}^{2}}$,
$m_{i}=\frac{m_{iD}}{m_{0}}$, $l=\frac{l_{D}}{l_{0}}$, $c_{\varphi}=\frac{c_{\varphi D}}{m_{0}l_{o}^{2}\omega_{0}}$,
$N_{0}=\frac{N_{0D}}{m_{0}l_{o}^{2}\omega_{0}^{2}}$, $N_{1}=\frac{N_{1D}}{m_{0}l_{o}^{2}\omega_{0}}$,
$M=\frac{M_{D}}{m_{0}}$, $g=\frac{g_{D}}{l_{o}\omega_{0}^{2}}$,
$c_{x}=\frac{c_{xD}}{m_{0}\omega_{0}}$ and dimensionless control
parameter $k_{x}=\frac{k_{xD}}{m_{0}\omega_{0}^{2}}$. Dimensionless
parameters have the following values: {$m_{i}=\frac{2.0}{n}$,
$l=0.25$, $c_{\varphi}=\frac{0.02}{n}$, $N_{0}=5.0$, $N_{1}=0.5$,
$M=6.0$, $g=9.81$.}

\end{document}